\documentclass[20pt]{article}
\usepackage{amsmath}
\usepackage{amssymb}
\usepackage{latexsym}
\usepackage{amsmath}
\usepackage{lineno}
\usepackage{graphicx}
\usepackage{float}
\usepackage{subfigure}
\usepackage{geometry}
\usepackage{graphics}
\usepackage{subfigure}
\usepackage{graphicx}
\usepackage{multirow}
\usepackage{booktabs}
\usepackage{hyperref}
\makeatletter

\newcommand{\be}{\begin{equation}}
\newcommand{\ee}{\end{equation}}

\begin{document}
\begin{center}
\large {\bf Quantum corrections to the  thermodynamics of Schwarzschild-Tangherlini black hole and the generalized uncertainty principle}
\end{center}

\begin{center}
Z. W. Feng$^{1}$  $\footnote{E-mail:  zwfengphy@163.com}$,
H. L. Li,$^{1,2}$
X. T. Zu,$^{1}$ $\footnote{E-mail: xtzu@uestc.edu.cn}$
S. Z. Yang$^{3}$ $\footnote{E-mail: szyangcwnu@163.com}$
\end{center}

\begin{center}
\textit{1.School of Physical Electronics, University of Electronic Science and Technology of China, Chengdu, 610054, China\\
2.College of Physics Science and Technology, Shenyang Normal University, Shenyang, 110034, China\\
3.Physics and Space Science College, China West Normal University, Nanchong, 637009, China}
\end{center}

\noindent
{\bf Abstract:} We investigate the thermodynamics of the Schwarzschild-Tangherlini black hole in the context of the generalized uncertainty  principle (GUP). The corrections to the Hawking temperature, entropy and the heat capacity are obtained via the modified Hamilton-Jacobi equation.
These modifications show that the GUP changes the evolution of the Schwarzschild-Tangherlini black hole.
Specially, the GUP effect becomes susceptible when the radius or mass of the black hole approaches the order of Planck scale,
it stops radiating and leads to black hole remnant. Meanwhile, the Planck scale remnant can be confirmed through the analysis of the heat capacity.
Those phenomena imply that the GUP may give a way to solve the information paradox. Besides, we also investigate the possibilities to observe
the black hole at  the Large Hadron Collider (LHC), and the results demonstrate that the black hole cannot be produced in the recent LHC.

\noindent
\section{ Introduction}
One common feature among various quantum gravity theories, such as string theory, loop quantum gravity and non-commutative geometry,
is the existence of a minimum measurable length which can be identified with the order of the Planck scale \cite{ch1,ch2,ch3,ch4}.
This view is also advocated by many Gedanken experiments \cite{ch12}.
The minimum measurable length is especially important since it can be applied into different physical systems and modify many
classical theories \cite{ch7a+,ch13,ch6,ch5,ch14,ch4+,ch5+,ch6+}. One of the most interesting modified theories is called the generalized uncertainty
principle (GUP), which is a generalization of the conventional Heisenberg uncertainty principle (HUP).
It is well known that the uncertainty principle is closely related to the fundamental commutation relation.
Therefore, taking account of the minimum measurable scale, Kempf, Mangano and Mann proposed a modified fundamental commutation relation
\be
\label{eq2}
\left[ {x_i ,p_j } \right] = i\hbar \delta _{ij} \left[ {1 + \beta p^2 } \right],
\ee
with the position and momentum operators
\be
\label{eq1+}
\begin{array}{*{20}c}
   {x_i  = x_{0i} ,} & {p_j  = p_{0j} \left( {1 + \beta p_0^2 } \right),}  \\
\end{array}
\ee
 where $x_{0i}$  and  $p_{0j} $ satisfy  the canonical commutation relations  $\left[ {x_{0i} ,p_{0j} } \right] = i\hbar \delta _{ij}$ \cite{ch11}.
 Through the above equations, the most studied form of the GUP is derived as
\begin{equation}
\label{eq1}
\Delta x\Delta p \ge \frac{\hbar }{2}\left[ {1 + \beta \left( {\Delta p} \right)^2 } \right],
\end{equation}
where $\Delta x$ and $\Delta p$ represent  the uncertainties for position and momentum.
The $\beta  = {{\beta _0 \ell_p^2 } \mathord{\left/ {\vphantom {{\beta _0 \ell_p^2 } {\hbar ^2 }}} \right. \kern-\nulldelimiterspace} {\hbar ^2 }} = {{\beta _0 }
\mathord{\left/ {\vphantom {{\beta _0 } {M_p^2 }}} \right. \kern-\nulldelimiterspace} {M_p^2 }}c^2$,   $\beta_0$ ($\leq 10^{34}$)
is a dimensionless constant, and  $\ell_p$ and $M_p$ are the Planck length $(\sim10^{-35}m)$ and Planck mass, respectively.
In the HUP framework, the position uncertainty can be measured to an arbitrary small value since there is no restriction on the measurement precision of momentum of the particles. However, Eq. (\ref{eq1}) implies the GUP existence of minimum measurable length  $\Delta x_{min} \approx \ell_{p} \sqrt{{\beta _{0}}} $.
In the limit $\Delta x \gg \ell_p$, one recovers the HUP $\Delta x\Delta p \ge {\hbar  \mathord{\left/  {\vphantom {\hbar  2}} \right. \kern-\nulldelimiterspace} 2}$.

The implications of the aspects of GUP have been investigated in many contexts such as modifications of quantum Hall effect \cite{ch15},
neutrino oscillations \cite{ch16}, Landau levels \cite{ch16c+} and cosmology \cite{ch17c+,ch17},  the weak equivalence principle (WEP) \cite{ch18f+}
and Newton's  law \cite{ch16+,ch16b+}. It should be noted that the GUP has also influence on the thermodynamics of black holes.
In an elegant paper, Adler, Chen and Santiago proposed that the $\Delta p$ and $\Delta x$ of the GUP can be identified as the temperature
and radius of the black hole. With this heuristic method (the Hawking temperature-uncertainty relation), the GUP's impacts on the thermodynamics of Schwarzschild (SC) black hole have been discussed in \cite{ch18}. This work showed that the modified Hawking temperature is higher than the original case,
and the GUP effect leads to the remnants in the final stages of black hole evaporation. This interesting work has widely got attention, many other black holes' thermodynamics have been studied with the help of the Hawking temperature-uncertainty relation \cite{ch17+,ch19,ch32b+,ch19+,ch21}.

On the other hand, the thermodynamics of black holes also can be calculated by the tunneling method \cite{ch26a+,ch27a+,ch28a+,ch1x+,ch29a+,ch30a+}.
The tunneling method was first proposed by Parikh and Wilczek for investigating the tunneling behaviors of massless scalar particles \cite{ch26a+}.
Later, this method was extended to a study of the tunneling of massive and charged scalar particles\cite{ch27a+}.
The Hamilton-Jacobi ansatz is another kind of tunneling method \cite{ch28a+,ch1x+,ch29a+}.
With the help of Hamilton-Jacobi ansatz, Kerner and Mann have carefully analyzed the fermion tunneling from black holes \cite{ch30a+}.
So far, the tunneling method plays an important role in studying the black hole radiation, it can effectively help people further understand the properties of black holes, gravity and quantum gravity \cite{ch1e+,ch2e+,ch3e+,ch4e+,ch5e+}.

Combining the GUP with the tunneling method, Nozari and Mehdipour studied the modified tunneling rate of the SC black hole \cite{ch1c}. Subsequently,
many more papers on the subject appeared, aiming to investigate the GUP corrected temperature of complicated spacetimes \cite{ch18+,ch27,ch28,ch28+,ch29+,ch20+,ch20b+}. However, as far as we know, those works are limited to low dimensional spacetimes. It is well known that the higher dimensional spacetimes include more physics information, moreover, one of the most exciting signatures is that people may detect the black holes in the large extra dimensions by using the Large Hadron Collider (LHC) and the Ultrahigh Energy
Cosmic Ray Air Showers (UECRAS) \cite{ch22,ch22+,ch23+,ch23,ch24,ch25,ch3x3}. In other words, the large extra dimensions have opened
up new doors of research in black holes and quantum gravity. Therefore, in this paper, we will investigate the GUP corrected
thermodynamics of Schwarzschild-Tangherlini (ST) black hole by using the quantum tunneling method.
The ST black hole is a typical higher dimensional black hole, people can get many new solutions of higher dimensional spacetimes via the ST metrics.
In \cite{ch26}, the authors showed that the ST black hole is a good approximation to a compactified spacetime when the compact dimension's size
is much larger than the black hole's size. To acquire a deeper understanding of Schwarzschild metric, one can study \cite{ch26xx+,ch27xx+}. Thus, the ST black hole is a good tool for researching the distorted compactified spacetime.
Based on the above arguments, we think the GUP corrected thermodynamics of ST black hole are worth to be studied.
 By utilizing the tunneling method and the GUP, we find that the modified temperature is lower than the original case.
 Meanwhile, it is also in contrast to the earlier findings, which are analyzed by the Hawking temperature-uncertainty relation \cite{ch18,ch23}.
 When the mass of the ST black hole reaches the order of the Planck scale, the GUP corrected thermodynamics decreases to zero.
 This in turn prevents black hole from evaporating completely and leads to a remnant of the ST black hole.

The outline of the paper is as follows. In Sect. 2, incorporating GUP, we derive the modified Hamilton-Jacobi equations in the curved spacetime via WKB
approximation. In Sect. 3, the tunneling radiation of particles from the ST black hole is addressed.
In Sect. 4, due to the GUP corrected temperature, we analyze the remnants of ST black hole. In Sect. 5, we investigate the minimum black hole energy to form a black hole in the LHC. The last section is devoted to our conclusion.

\section{Modified Hamilton-Jacobi equations}
In this section, we will derive the modified Hamilton-Jacobi equations from the generalized Klein-Gordon equation and  the  generalized Dirac equation. Based on momentum operators of Eq. (\ref{eq1+}), the square of the momentum takes the form \cite{ch27,ch28}
\begin{equation}
\label{eq3}
p^2  = p_i p^i  \simeq  - \hbar ^2 \left[ {1 - 2\beta \hbar ^2 \left( {\partial _j \partial ^j } \right)} \right]\left( {\partial _i \partial ^i } \right).
\end{equation}
It is  noted that the higher-order  terms $\mathcal{O}(\beta)$  in above equation are ignored.
Adopting the effects of generalized frequency $\bar \omega  = E\left( {1 - \beta E^2 } \right)$  and the mass shell condition,
the generalized expression of the energy is \cite{ch29}
\begin{equation}
\label{eq4}
\bar E = E\left[ {1 - \beta \left( {p^2  + m^2 } \right)} \right],
\end{equation}
where energy operator is defined as  $E = i\hbar \partial _t$. Therefore, the original Klein-Gordon equation in the curved spacetime is given by
\begin{equation}
\label{eq5}
\left[ {\left( {i\hbar } \right)^2 D^\mu  D_\mu   + m^2 } \right]\Psi  = 0,
\end{equation}
where  $D_\mu =\nabla _\mu   + {{ieA_\mu  } \mathord{\left/ {\vphantom {{ieA_\mu  } \hbar }} \right. \kern-\nulldelimiterspace} \hbar }$ with the geometrically covariant derivative $\nabla _\mu$; the $m$  and $e$  denote the mass and charge of the particles, $A^\mu$  is the electromagnetic potential of spacetime. In order to get the generalized Klein-Gordon equation, Eq. (\ref{eq5}) should be rewritten as
\begin{equation}
\label{eq6}
{ - \left( {i\hbar } \right)^2 \left( {\partial _t  + \frac{i}{\hbar }eA_t } \right)\left( {\partial ^t  + \frac{i}{\hbar }eA^t } \right)\Psi  = \left[ {\left( {i\hbar } \right)^2 \left( {\partial _k  + \frac{i}{\hbar }eA_k } \right)\left( {\partial ^k  + \frac{i}{\hbar }eA^k } \right) + m^2 } \right]\Psi ,}
\end{equation}
where $k = 1,2,3 \cdots$ represent the spatial coordinates. In the above equation, the relation $\nabla _\mu   = \partial _\mu $ has been used. The right hand of Eq. (\ref{eq6}) is related to the energy. Inserting the Eqs. (\ref{eq3}) and (\ref{eq4}) into above equation, one can generalize the original Klein-Gordon equation to the following form:
\begin{equation}
\label{eq7}
{ - \left( {i\hbar } \right)^2 \left( {\partial _t  + \frac{i}{\hbar }eA_t } \right)\left( {\partial ^t  + \frac{i}{\hbar }eA^t } \right)\Psi  = \left[ {\left( {i\hbar } \right)^2 \left( {\partial _k  + \frac{i}{\hbar }eA_k } \right)\left( {\partial ^k  + \frac{i}{\hbar }eA^k } \right) + m^2 } \right]\left[ {1 - \beta \left( {p^2  + m^2 } \right)} \right]^2 \Psi .}
\end{equation}
 The wave function of generalized Klein-Gordon equation Eq. (\ref{eq7}) can be expressed as  $\Psi  = \exp [{{iS(t,k)} \mathord{\left/ {\vphantom {{iS(t,k)} \hbar }} \right. \kern-\nulldelimiterspace} \hbar }]$, where $S\left( {t,k}\right)$ is the action of the scalar particle. Substituting the wave function into Eq. (\ref{eq7}) and using the WKB approximation, the modified Hamilton-Jacobi equation for the scalar particle is got  as
\begin{equation}
\label{eq8}
{g^{00} \left( {\partial _0 S + eA_0 } \right)^2  + \left[ {g^{kk} \left( {\partial _k S + eA_k } \right)^2  + m^2 } \right]\left\{ {1 - 2\beta \left[ {g^{jj} \left( {\partial _j S} \right)^2  + m^2 } \right]} \right\} = 0.}
\end{equation}

It is well known that the original Dirac equation can be expressed as
 $ - i\gamma ^t \nabla _t \Psi  = (i\gamma ^k \nabla _k  + {m \mathord{\left/  {\vphantom {m \hbar }} \right.  \kern-\nulldelimiterspace} \hbar })\Psi$
 with $\nabla _k   = \partial _k   + \Omega _k   + {{ieA_k  } \mathord{\left/ {\vphantom {{ieA_k  } \hbar }} \right. \kern-\nulldelimiterspace} \hbar }$,
 where the left hand is related to the energy. According to the method in \cite{ch27,ch28},
putting the generalized expression of  the energy,  Eqs. (\ref{eq3}) and (\ref{eq4}) into the original Dirac equation,
one  finds the generalized Dirac equation in curved spacetime,
\begin{equation}
 - i\gamma ^t \nabla _t \Psi  = \left( {i\gamma ^k \nabla _k  + {m \mathord{\left/
 {\vphantom {m \hbar }} \right.
 \kern-\nulldelimiterspace} \hbar }} \right)\Upsilon \left( \beta  \right)\Psi,
 \label{eq9}
\end{equation}
where $\Upsilon \left( \beta  \right) = 1 - \beta \left( {p^2  + m^2 } \right)$. Since the $t-t$ component of Eq. (\ref{eq9}) is related to the energy, it did not get corrected by the GUP term $\Upsilon \left( \beta  \right)$, thus the Eq. (\ref{eq9}) is different from the generalized Dirac equation $ - i\gamma ^0 \partial _0 \Psi  = (i\gamma ^i \nabla _i  + i\gamma ^t \Omega _t  + {{ieA_t } \mathord{\left/  {\vphantom {{ieA_t } \hbar }} \right.  \kern-\nulldelimiterspace} \hbar } + {m \mathord{\left/  {\vphantom {m \hbar }} \right.  \kern-\nulldelimiterspace} \hbar })\Upsilon \left( \beta  \right)\Psi $ in \cite{ch27,ch28}. Then, multiplying $ - i\gamma ^t \nabla _t  - [i\gamma ^n \nabla _n  - {m \mathord{\left/ {\vphantom {m \hbar }} \right.
 \kern-\nulldelimiterspace} \hbar }]\Upsilon \left( \beta  \right)$ by Eq. (\ref{eq9}), the generalized Dirac equation can be written as
\begin{equation}
\label{eq10}
\begin{array}{l}
 \left\{ { - \left( {\gamma ^t \nabla _t } \right)^2  - \gamma ^t \nabla _t \gamma ^n \nabla _n \Upsilon \left( \beta  \right) - \gamma ^k \nabla _k \gamma ^t \nabla _t \Upsilon \left( \beta  \right)} \right. +  \left. {\left[ {i\left( {\gamma ^n \nabla _n  - \gamma ^k \nabla _k } \right)\frac{m}{\hbar } - \gamma ^k \nabla _k \gamma ^n \nabla _n  - \left( {\frac{m}{\hbar }} \right)^2 } \right]\Upsilon \left( \beta  \right)^2 } \right\} \\
 \\
  \times \Psi  = 0. \\
 \end{array}
\end{equation}
Assuming  $k=n$, the above equation becomes to
\begin{equation}
\label{eq11}
{\left\{ { - \frac{{\left\{ {\gamma ^t ,\gamma ^t } \right\}}}{2}\nabla _t^2  - \left[ {\frac{{\left\{ {\gamma ^k ,\gamma ^k } \right\}}}{2}\nabla _k^2  + \left( {\frac{m}{\hbar }} \right)^2 } \right]\Upsilon \left( \beta  \right)^2 } \right\}\Psi  = 0.}
\end{equation}
In order to derive the modified Hamilton-Jacobi equation from Eq. (\ref{eq11}), the wave function of generalized Dirac equation takes on the form
\begin{equation}
\label{eq12}
\Psi  = \xi \left( {t,k} \right)\exp \left[ {{{iS\left( {t,k} \right)} \mathord{\left/
 {\vphantom {{iS\left( {t,k} \right)} \hbar }} \right.
 \kern-\nulldelimiterspace} \hbar }} \right],
\end{equation}
where  $\xi \left( {t,k} \right)$ is a vector function of the spacetime. Denoting  $t=0$, the gamma matrices'  anti-commutation  relations obey $\left\{ {\gamma ^0 ,\gamma ^k } \right\} = 0$ ,  $\left\{ {\gamma ^k ,\gamma ^k } \right\} = 2g^{kk} I$, and $\left\{ {\gamma ^0 ,\gamma ^0 } \right\} = 2g^{00} I$. Substituting the gamma  matrices' anti-commutation relations and Eq. (\ref{eq12}) into Eq. (\ref{eq11}), the resulting equation to leading order in $\beta$ is
\begin{equation}
\label{eq15+}
{\left\{ {g^{00} \left( {\partial _t S + eA_t } \right)^2  + \left[ {g^{kk} \left( {\partial _k S + eA_k } \right)^2  + m^2 } \right]\left. {\left\{ {1 - 2\beta \left[ {g^{jj} \left( {\partial _j S} \right)^2  + m^2 } \right]} \right\}} \right\}\xi \left( {t,k} \right) = 0.} \right.}
\end{equation}
Equation  (\ref{eq15+}) for the coefficient will has a non-trivial solution if and only if the determinant vanishes, that is
\begin{equation}
\label{eq15}
{{\rm{Det}}\left\{ {g^{00} \left( {\partial _t S - eA_t } \right)^2  + \left[ {g^{kk} \left( {\partial _k S + eA_k } \right)^2  + m^2 } \right]\left\{ {1 - 2\beta \left[ {g^{jj} \left( {\partial _j S} \right)^2  + m^2 } \right]} \right\}} \right\} = 0.}
\end{equation}
When keeping the leading-order term of $\beta$, the modified Hamilton-Jacobi equation for fermion is directly obtained:
\begin{equation}
\label{eq16}
{g^{00} \left( {\partial _0 S + eA_0 } \right)^2  + \left[ {g^{kk} \left( {\partial _k S + eA_k } \right)^2  + m^2 } \right]\left\{ {1 - 2\beta \left[ {g^{jj} \left( {\partial _j S} \right)^2  + m^2 } \right]} \right\} = 0.}
\end{equation}
Comparing Eq. (\ref{eq8}) with Eq. (\ref{eq16}), it is clear that the modified Hamilton-Jacobi equations for a scalar particle and fermions are the similar. In \cite{ch3e+,ch20+,ch31d+}, the authors derived the Hamilton-Jacobi from the Rarita-Schwinger equation, the Maxwell equations and the gravitational wave equation, they indicated that the Hamilton-Jacobi equation can describe the behavior of particles with any spin in the curve spacetime.
As is well known, the Hamilton-Jacobi ansatz can greatly simplify the workload in the research of black hole radiation.
Especially for the fermion tunneling case, we do not need to construct the tetrads and gamma matrices with the help of the Hamilton-Jacobi equation.
Adopting the modified Hamilton-Jacobi equation, the tunneling radiation of ST black hole will be studied in the next section.

\section{Quantum tunneling from ST black hole}
To begin with, we need make a few remarks about the ST black hole. In Ref.~\cite{ch32}, the author added extra compact spatial dimensions to a static spherically symmetric spacetime, and obtained the line element of the ST black hole
\begin{equation}
\label{eq17}
ds^2  =  - f\left( r \right)dt^2  + f\left( r \right)^{ - 1} dr^2  + r^2 d\Omega _{D - 2}^2 ,
\end{equation}
where $f\left( r \right) = 1 - \left( {{{r_H } \mathord{\left/ {\vphantom {{r_H } r}} \right. \kern-\nulldelimiterspace} r}} \right)^{D - 3}$,  $d\Omega _{D - 2}^2$ is the metric on a unit $D-2$  dimensional sphere, it covered by the original angular coordinates  $\theta _1 ,\theta _2 ,\theta _{3}, \cdots, \theta _{D - 2}$.  $r_H$ is the event horizon of the ST black hole, which is characterized by the mass $M$
\begin{equation}
\label{eq18}
r_H  = \left[ {\frac{{16GM}}{{\left( {D - 2} \right)\varpi _{D - 2} }}} \right]^{\frac{1}{{D - 3}}}  = \frac{1}{{\sqrt \pi  }}\left[ {\frac{{8M\Gamma \left( {\frac{{D - 1}}{2}} \right)}}{{D - 2M_P^{D - 2} }}} \right]^{\frac{1}{{D - 3}}}
\end{equation}
where $G = {1 \mathord{\left/ {\vphantom {1 {M_P^{D - 2} }}} \right. \kern-\nulldelimiterspace} {M_P^{D - 2} }}$ is the $D-2$ dimensional Newton constant  and the volume of the unit $D-2$ dimensional sphere as  $\varpi _{D - 2}  = {{2\pi ^{\frac{{D - 1}}{2}} } \mathord{\left/ {\vphantom {{2\pi ^{\frac{{D - 1}}{2}} } {\Gamma \left( {\frac{{D - 1}}{2}} \right)}}} \right.  \kern-\nulldelimiterspace} {\Gamma \left( {\frac{{D - 1}}{2}} \right)}}$ \cite{ch22}.

Next, we will calculate the quantum tunneling from the ST black hole. Inserting the inverse metric of ST black hole into  the modified Hamilton-Jacobi equation, one has
\begin{equation}
\label{eq20}
\begin{array}{l}
 f^{ - 1} \left( {\partial _t S} \right)^2  - \left[ {f\left( r \right)\left( {\partial _r S} \right)^2  + \left( {g^{\theta _1 \theta _1 } } \right)\left( {\partial _{\theta _1 } S} \right)^2  + \left( {g^{\theta _2 \theta _2 } } \right)\left( {\partial _{\theta _2 } S} \right)^2  +  \cdots  + \left( {g^{\theta _{D - 2} \theta _{D - 2} } } \right)\left( {\partial _{\theta _{D - 2} } S} \right)^2 } \right. \\
 \\
 \left. { + m^2 } \right]\left\{ {1 - 2\beta \left[ {f\left( r \right)\left( {\partial _r S} \right)^2  + \left( {g^{\theta _1 \theta _1 } } \right)\left( {\partial _{\theta _1 } S} \right)^2  + \left( {g^{\theta _2 \theta _2 } } \right)\left( {\partial _{\theta _2 } S} \right)^2  +  \cdots  + \left( {g^{\theta _{D - 2} \theta _{D - 2} } } \right)\left( {\partial _{\theta _{D - 2} } S} \right)^2 } \right.} \right. \\
 \\
 \left. {\left. { + m^2 } \right]} \right\} = 0. \\
 \end{array}
\end{equation}
Since the spacetime of ST black hole is static, the action $S$ is supposed to take the form \(S = - \omega
t  + W\left(  r  \right)  +  \Theta  \left(  {\theta_1  ,\theta_2 ,\ldots  ,\theta_{D -  2}  }  \right)  \), where \(\omega \) is the energy of the emitted particles. Equation (\ref{eq20}) can be written as
\begin{equation}
\label{eq21}
\begin{array}{*{20}c}
   {f\left( r \right)\left( {\partial _r W} \right)^2 \left[ {2\beta f^{ - 1} \left( r \right)\left( {\partial _r W} \right)^2  - 1} \right] + \left[ {f\left( r \right)\left( {\partial _r W} \right)^2  + \left( {g^{\theta _1 \theta _1 } } \right)\left( {\partial _{\theta _1 } \Theta } \right)^2 } \right. + \left( {g^{\theta _2 \theta _2 } } \right)}  \\
   {}  \\
   {\left. { \times \left( {\partial _{\theta _2 } \Theta } \right)^2  +  \cdots  + \left( {g^{\theta _{D - 2} \theta _{D - 2} } } \right)\left( {\partial _{\theta _{D - 2} } S} \right)^2 } \right] \left[ {4\beta f^{ - 1} \left( r \right)\left( {\partial _r W} \right)^2  - 1} \right] + \omega ^2 f^{ - 1} \left( r \right) =  - \lambda ,}  \\
\end{array}
\end{equation}
\begin{equation}
\label{eq22}
{{\rm{2}}\beta \left[ {\left( {g^{\theta _1 \theta _1 } } \right)\left( {\partial _{\theta _1 } \Theta } \right)^2  + \left( {g^{\theta _2 \theta _2 } } \right)\left( {\partial _{\theta _2 } \Theta } \right)^2 \left. { +  \cdots  + \left( {g^{\theta _{D - 2} \theta _{D - 2} } } \right)\left( {\partial _{\theta _{D - 2} } \Theta } \right)^2 } \right]^{\rm{2}} {\rm{ = }}\lambda .} \right.}
\end{equation}
where $\lambda$ is a constant. First, focusing on Eq. (\ref{eq22}), in \cite{ch20+}, the author showed that the magnitude of the  particles' angular momentum  can be expressed in the terms of $\partial _{\theta _1 } \Theta$, $\partial _{\theta _2 } \Theta$,$ \cdots \partial _{\theta _{D - 2} } \Theta$, that is
\begin{equation}
\label{eq22k}
 \left( {g^{\theta _1 \theta _1 } } \right)\left( {\partial _{\theta _1 } \Theta } \right)^2  + \left( {g^{\theta _2 \theta _2 } } \right)\left( {\partial _{\theta _2 } \Theta } \right)^2  + \cdots  + \left( {g^{\theta _{D - 2} \theta _{D - 2} } } \right)\left( {\partial _{D - 2} \Theta } \right)^2  = \mathcal{L}^2,  \\
\end{equation}
According to Eq. (\ref{eq22k}), one can write Eq. (\ref{eq22}) as
\begin{equation}
\label{eq23}
2\left( {{\cal L}^2 } \right)^2  = {\lambda  \mathord{\left/
 {\vphantom {\lambda  \beta }} \right.
 \kern-\nulldelimiterspace} \beta }.
\end{equation}
In the above equation  is  indicated  that the constant $\lambda$ is related to the angular momentum of the emitted particle.  With the help of Eqs. (\ref{eq22k}) and (\ref{eq23}), Eq. (\ref{eq21}) becomes
\begin{equation}
\label{eq24}
P_4 \left( {\partial _r W} \right)^4  + P_2 \left( {\partial _r W} \right)^2  + P_0  = 0,
\end{equation}
where   $P_4  = 2\beta f\left( r \right)^2$,   $P_2  = \left( {4m^2 \beta  - 1} \right)f\left( r \right)$ and   $P_0  = \omega ^2 f^{ - 1} \left( r \right) + \left( {2m^2 \beta  - 1} \right)m^2$. Neglecting the higher orders $\beta$ of and solving above equation, one finds

\begin{equation}
\label{eq25}
W_ \pm   = \pm \frac{1}{{f\left( r \right)}}\sqrt {f\left( r \right)\left( {m^2  - \lambda  + \sqrt {{\lambda  \mathord{\left/
 {\vphantom {\lambda  {2\beta }}} \right.
 \kern-\nulldelimiterspace} {2\beta }}} } \right) + \omega ^2 }   \left\{ {1 + \beta \left[ {m^2  + f^{ - 1} \left( r \right)\omega ^2 } \right] + \sqrt {{{\beta \lambda } \mathord{\left/
 {\vphantom {{\beta \lambda } 2}} \right.
 \kern-\nulldelimiterspace} 2}} } \right\}dr,
\end{equation}
where the ${ +  \mathord{\left/ {\vphantom { +   - }} \right. \kern-\nulldelimiterspace}  - }$ denote the outgoing/incoming solutions of the emitted particles. In order to solve above equation, one needs to find the residue of Eq. (\ref{eq25}) on the event horizon. By expanding a Laurent series on the event horizon and keeping the first-order term of  $\beta$, the result of Eq. (\ref{eq25}) takes on the form as
\begin{equation}
\label{eq26}
 W\left( {r_H } \right)_ \pm   =  \pm \frac{{i\pi r_H \omega }}{{D - 3}}\left\{ {1 + \sqrt {\frac{{\beta \lambda }}{8}}  + } \right.\beta  \left. {\left[ {\frac{{m^2  + \lambda }}{2} + \frac{{\left( {D - 2} \right)\omega ^2 }}{{\left( {D - 3} \right)}}} \right]} \right\} + \Delta \left( {realpart} \right).
\end{equation}
Because the real part of Eq. (\ref{eq26}) is irrelevant to the tunneling rate, we only keep the imaginary part. For obtaining the tunneling rate from
Eq. (\ref{eq26}), one needs to solve the factor-two problem \cite{ch33,chvv1+}.
One of the best ways to solve this problem is to adopt the temporal contribution expression.
According to \cite{chv1+,chv2+,chv3+,chv4+,chv5+}, the spatial part of the tunneling rate of emitted particle is
\begin{eqnarray}
 \Gamma & \propto & \exp \left( { - {\mathop{\rm Im}\nolimits} \oint {p_r dr} } \right)  =  \exp \left[ {{\mathop{\rm Im}\nolimits} \left( {\int {p_r^{out} dr}  - \int {p_r^{in} dr} } \right)} \right]
 \nonumber \\
 \nonumber \\
 & = & \exp \left\{ { - \frac{{2\pi r_H \omega }}{{D - 3}}\left\{ {1 + \sqrt {\frac{{\beta \lambda }}{8}}  + \beta \left[ {\frac{{m^2  + \lambda }}{2} + \frac{{\left( {D - 2} \right)\omega ^2 }}{{\left( {D - 3} \right)}}} \right]} \right\}} \right\},
\label{eq1v+}
\end{eqnarray}
where $p_r  = \partial _r W$. However, as pointed out in \cite{chv2+}, the authors showed that the temporal contribution to the tunneling amplitude was lost in the above discussion. For incorporating the temporal contribution into our calculation, we need use Kruskal coordinates $(T,R)$. The  exterior region is given by
\begin{equation}
\label{eq2v+}
\begin{array}{*{20}c}
   {T = \exp \left( {\kappa r_* } \right)\sinh \left( {\kappa t} \right),} & {R = \exp \left( {\kappa r_* } \right)\cosh \left( {\kappa t} \right)}  \\
\end{array},
\end{equation}
where $r_*  = r + \frac{1}{{2\kappa }}\ln \frac{{r - r_H }}{{r_H }}$ is the tortoise coordinate and $\kappa$ is the surface gravity  of the ST black hole. In order to connect the interior region and the exterior region across the horizon, one can rotate the time $t$ as $t\rightarrow t - {{i\pi } \mathord{\left/ {\vphantom {{i\pi } {2\kappa }}} \right. \kern-\nulldelimiterspace} {2\kappa }}$. By this operation, one obtains an additional imaginary contribution ${\mathop{\rm Im}\nolimits} \left( {\omega \Delta t^{out,in} } \right) = \omega {\pi  \mathord{\left/ {\vphantom {\pi  {2\kappa }}} \right.  \kern-\nulldelimiterspace} {2\kappa }}$. Therefore, the total temporal contribution becomes to ${\mathop{\rm Im}\nolimits} \omega \Delta t = \omega {\pi  \mathord{\left/ {\vphantom {\pi  \kappa }} \right. \kern-\nulldelimiterspace} \kappa }$. According to Eq. (\ref{eq1v+}), the GUP corrected tunneling rate of emitted particle across the horizon is derived to be
\begin{equation}
\label{eq281}
 \Gamma  \propto \exp \left[ { - {\mathop{\rm Im}\nolimits} \left( {\omega t + {\mathop{\rm Im}\nolimits} \oint {p_r dr} } \right)} \right]
 = \exp \left\{ { - \frac{{4\pi r_H \omega }}{{D - 3}}\left\{ {1 + \sqrt {\frac{{\beta \lambda }}{8}}  + \beta \left[ {\frac{{m^2  + \lambda }}{2} + \frac{{\left( {D - 2} \right)\omega ^2 }}{{\left( {D - 3} \right)}}} \right]} \right\}} \right\},
\end{equation}
Employing the Boltzmann factor, the GUP corrected Hawking temperature is
\begin{equation}
\label{eq28}
T_H  = T_0 \left\{ {1 + \sqrt {\frac{{\beta \lambda }}{8}}  + \beta \left[ {\frac{1}{2}\left( {m^2  + \lambda } \right) + \frac{{\left( {D - 2} \right)\omega ^2 }}{{\left( {D - 3} \right)}}} \right]} \right\}^{ - 1} ,
\end{equation}
where  $T_0  = {{\left( {D - 3} \right)} \mathord{\left/ {\vphantom {{\left( {D - 3} \right)} {4\pi r_H }}} \right. \kern-\nulldelimiterspace} {4\pi r_H }}$  is the semi-classical Hawking temperature of the ST black hole. Now, we turn to the  calculation of  the entropy of the ST black hole. Base on the first law of black hole thermodynamics, the entropy can be expressed as
\begin{equation}
\label{eq29}
 S = \int {T_H^{ - 1} dM}  = \int {\frac{{4\pi }}{{D - 3}}\left[ {\frac{{\left( {D - 2} \right)\varpi }}{{16\pi GM}}} \right]^{\frac{1}{{D - 3}}} }  \left\{ {1 + \sqrt {\frac{{\beta \lambda }}{8}}  + \beta \left[ {\frac{1}{2}\left( {m^2  + \lambda } \right) + \frac{{\left( {D - 2} \right)\omega ^2 }}{{\left( {D - 3} \right)}}} \right]} \right\}dM.
\end{equation}
The above equation cannot be evaluated exactly for general  $D$. According to the standard Hawing radiation theory, all particles near the event horizon seem effectively massless. Therefore, we do not consider the mass of the emitted particles in the following discussion.

\section{Remnants of ST black hole}
\begin{figure}
\centering
\subfigure{
\begin{minipage}[b]{0.3\textwidth}
\includegraphics[width=1\textwidth]{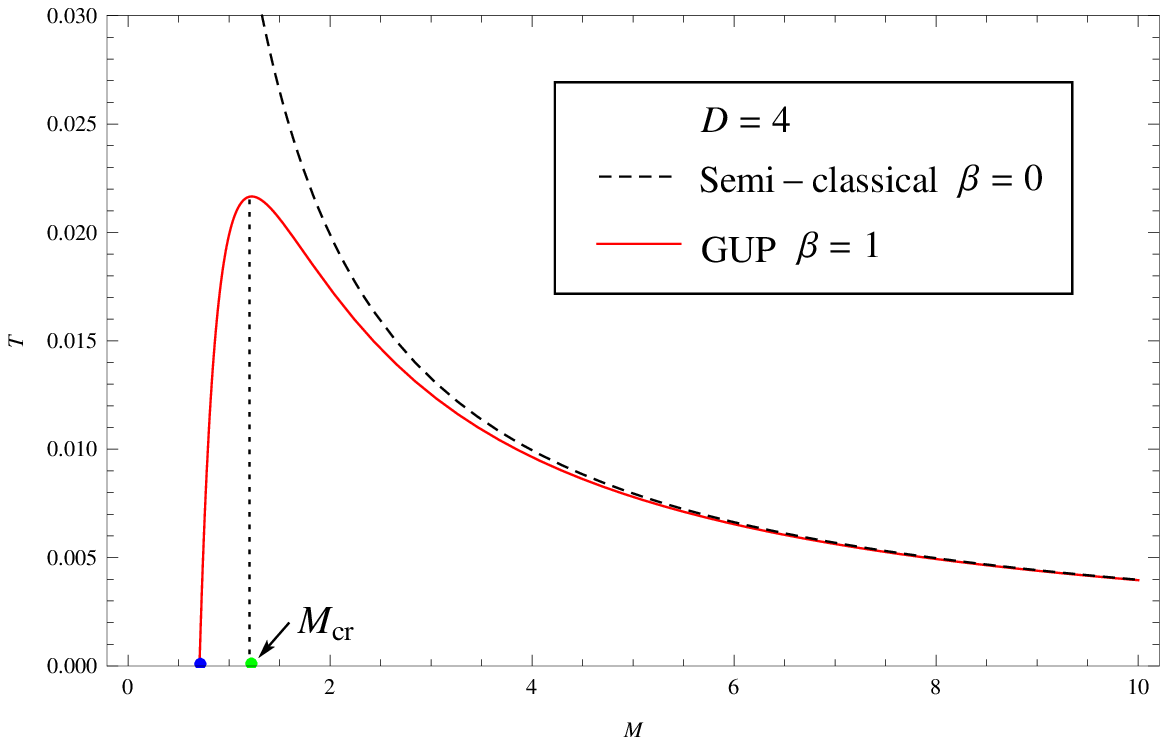}
\end{minipage}
}
\subfigure{
\begin{minipage}[b]{0.3\textwidth}
\includegraphics[width=1\textwidth]{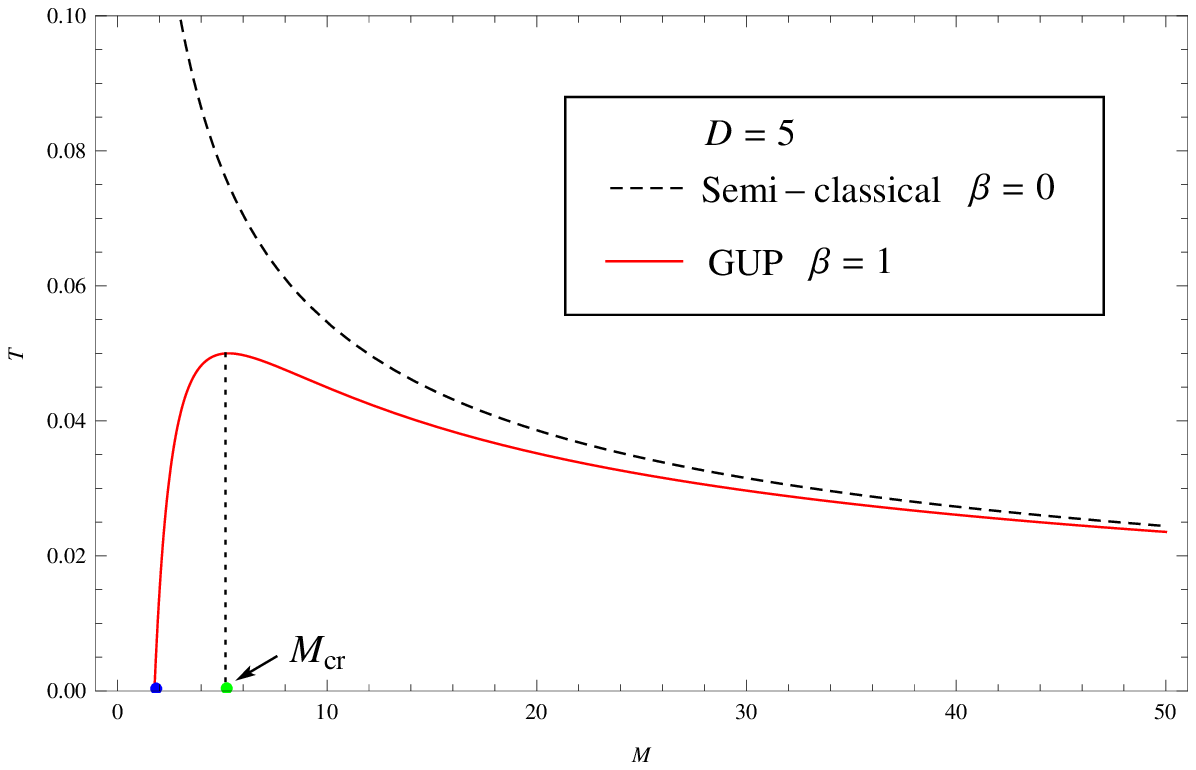}
\end{minipage}
}
\subfigure{
\begin{minipage}[b]{0.3\textwidth}
\includegraphics[width=1\textwidth]{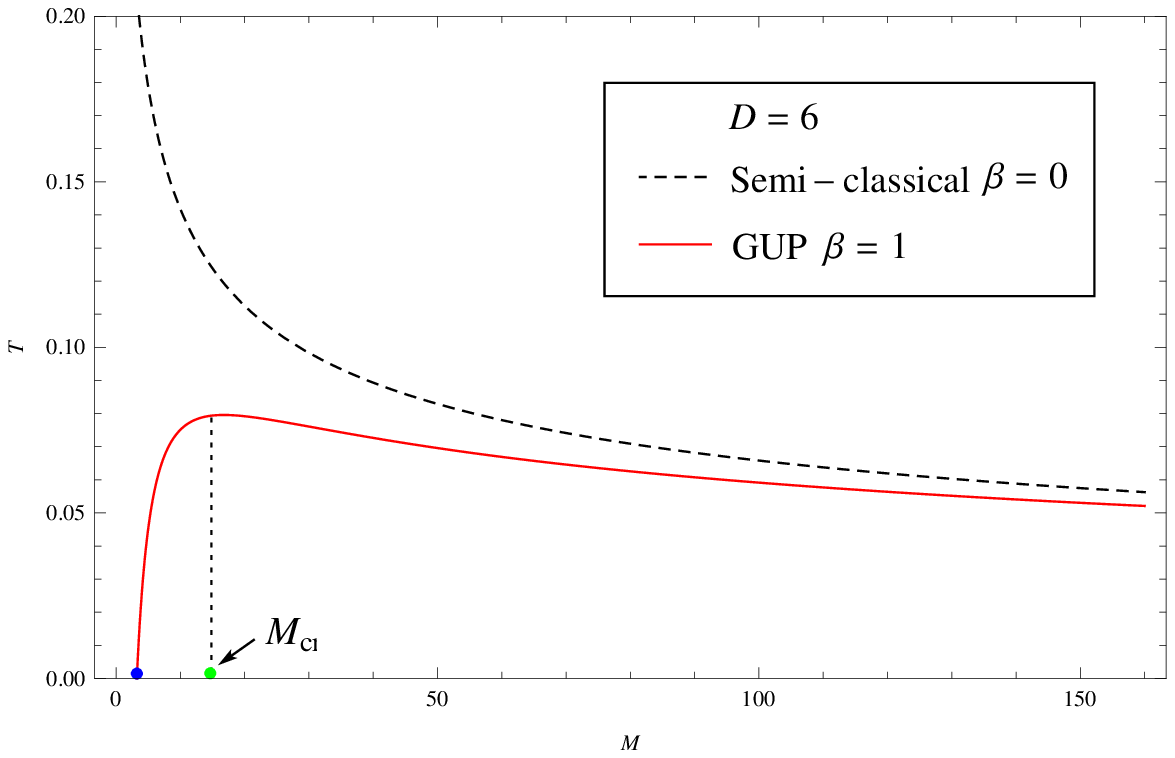}
\end{minipage}
}
\caption{Semi-classical and GUP corrected Hawking temperature of ST black hole for different values of mass. We set $M_P  = c = \hbar = 1$, $D = 4,5$ and $6$  for simplicity.}
\label{fig1}       
\end{figure}

A lot of work showed that the GUP can lead to a black hole remnant \cite{ch17+,ch19,ch32b+,ch19+,ch21,ch1c,ch18+,ch27,ch28,ch28+,ch29+,ch20+,ch20b+}.
Therefore, it is interesting to investigate the remnant of the ST black hole. According to the saturated form of the uncertainty principle,
one gets a lower bound on the energy of the emitted particle in Hawking radiation, which can be expressed as \cite{ch18,ch34}
\begin{equation}
\label{eq30}
\omega  \ge {\hbar  \mathord{\left/
 {\vphantom {\hbar  {\Delta x}}} \right.
 \kern-\nulldelimiterspace} {\Delta x}}.
\end{equation}
 Near the event horizon of the ST black hole, it is possible to take the value of the uncertainty in position as the radius of the black hole, that is \cite{ch24,ch25}
\begin{equation}
\label{eq31}
\Delta x \approx r_{BH} =  r_H.
\end{equation}
Putting Eqs. (\ref{eq30}) and (\ref{eq31}) into Eq. (\ref{eq28}), and expanding, one has
\begin{eqnarray}
\label{eq32}
 T_H & =& T_0 \left\{ {1 + \frac{3}{2}\sqrt {\frac{{\beta \lambda }}{2}}  + \beta \left[ {\frac{{\left( {D - 2} \right)\omega ^2 }}{{\left( {D - 3} \right)}} - \frac{\lambda }{2}} \right]} \right\}^{ - 1}
 \nonumber \\
  & \simeq & T_0 \left\{ {\frac{{2\left[ {4\left( {D - 2} \right)\hbar ^2 \beta  + \left( {D - 3} \right)r_H^2 \left( {\sqrt {2\beta \lambda }  + 2\beta \lambda  - 4} \right)} \right]}}{{r_H^2 \left( {D - 3} \right)\left( {\beta \lambda  - 8} \right)}}} \right\}.
\end{eqnarray}
It is clear that  $T_{H}$ sensitively depends  on the event horizon of the ST black hole, the spacetime dimension $D$, the angular momentum of the emitted particles and the quantum gravity effect  $\beta$. An important relation should be mentioned,
 when $r_H  < \sqrt {\frac{{4\left( {D - 2} \right)\beta \hbar ^2 }}{{\left( {D - 3} \right)\left( {4 - 2\beta \lambda  - \sqrt {2\beta \lambda } } \right)}}}$,
 the Hawking temperature goes to  negative  values,  and  it violates the laws of black hole thermodynamics and has no physical meaning.
  Therefore, this relation indicates the existence of a minimum radius, where the Hawking temperature equals zero,
   that is,
\begin{equation}
\label{eq33xx}
r_{\min }  = \sqrt {\frac{{4\left( {D - 2} \right)\beta \hbar ^2 }}{{\left( {D - 3} \right)\left( {4 - 2\beta \lambda  - \sqrt {2\beta \lambda } } \right)}}} = \ell _p \sqrt {\frac{{4\hbar ^2 \left( {D - 2} \right)\beta _0 }}{{\left( {D - 3} \right)\left( {4\hbar ^2  - 2\lambda \beta _0 \ell _p^2  - \ell _p \hbar \sqrt {2\lambda \beta _0 } } \right)}}}.
\end{equation}
In addition, we can also express  Eq. (\ref{eq32}) in terms of the mass of the ST black hole to obtain the temperature-mass relation
\begin{equation}
\label{eq33}
T_H  \simeq \frac{{D - 3}}{{4\pi }}\left[ {\frac{{\left( {D - 2} \right)\varpi _{D - 2} }}{{16\pi GM}}} \right]^{\frac{1}{{D - 3}}} {\frac{{2\left\{ {4\hbar ^2 \beta \left( {D - 2} \right) - \left( {D - 3} \right)\left[ {\frac{{\left( {D - 2} \right)\varpi _{D - 2} }}{{16\pi GM}}} \right]^{\frac{2}{{D - 3}}} \left( {4 - \sqrt {2\beta \lambda }  - 2\beta \lambda } \right)} \right\}}}{{\left( {D - 3} \right)\left[ {\frac{{\left( {D - 2} \right)\varpi _{D - 2} }}{{16\pi GM}}} \right]^{\frac{2}{{D - 3}}} \left( {\beta \lambda  - 8} \right)}}} .
\end{equation}
From Eq. (\ref{eq33}), we find  that the GUP corrected temperature has physical meaning as far as the mass of ST black hole satisfies the inequality $M \ge \frac{{\left( {D - 2} \right)\varpi _{D - 2} }}{{16\pi G}}[\frac{{4\left( {D - 2} \right)\hbar ^2 \beta }}{{\left( {D - 3} \right)\left( {4 - 2\beta \lambda  - \sqrt {2\beta \lambda } } \right)}}]^{\frac{{D - 3}}{2}}$, which implies that the mass of ST black hole has a minimum value,
\begin{eqnarray}
M_{\min } & = & \frac{{\left( {D - 2} \right)\varpi _{D - 2} }}{{16\pi G}}\left[ {\frac{{4\left( {D - 2} \right)\hbar ^2 \beta }}{{\left( {D - 3} \right)\left( {4 - 2\beta \lambda  - \sqrt {2\beta \lambda } } \right)}}} \right]^{\frac{{D - 3}}{2}}
 \nonumber \\
 \nonumber \\
 & = & \frac{{\left( {D - 2} \right)M_p }}{{8\Gamma \left( {\frac{{D - 1}}{2}} \right)}}\left[ {\frac{{4\pi \beta _0 \hbar ^2 \left( {D - 2} \right)}}{{c^2 \left( {D - 3} \right)\left( {4 - \frac{{2\lambda \beta _0 }}{{M_p^2 c^2 }} - \sqrt {\frac{{2\lambda \beta _0 }}{{M_p^2 c^2 }}} } \right)}}} \right]^{\frac{{D - 3}}{2}}
\label{eq34}
\end{eqnarray}
Obviously, the minimum mass is related to the  Planck mass. According to Eq. (\ref{eq33}) and Eq. (\ref{eq34}), the behaviors of GUP corrected Hawking temperature and the original Hawking temperature of ST black hole are plotted in Fig. \ref{fig1}.

In Fig. \ref{fig1}, the dashed  black lines and solid red lines in the diagrams illustrate the original Hawking temperature and the GUP corrected temperature of the ST black hole. It is easy to see that the GUP corrected temperature is lower than the original Hawking temperature. Besides, different values of  $D$ give similar behavior of the Hawking temperature. For a large mass of the black hole, the GUP corrected temperature tends to the
original  value  of  the Hawking temperature because the effect of quantum gravity is negligible at that scale. However, as the mass of the black hole decreases, the GUP corrected temperature reaches the maximum value (at the critical mass  $M_{cr}$, which is marked by a green dot), and then decreases to zero when the mass approaches the minimum value of the mass ($M_{min}\sim M _p$ , which is marked by a blue dot). The GUP corrected temperature is unphysical below the $M_{min}$, it signals the existence of a black hole remnant $M_{res}=M_{min}$. The black hole remnant can be further confirmed from the heat capacity.

Since the thermodynamic stability of black hole is determined by the heat capacity $\mathcal{C}$,
a further inspection of the existence of the black hole remnant can be made by investigating the heat capacity of the ST black hole. The GUP corrected heat capacity is given by
\begin{equation}
\label{eq35}
\mathcal{C }= T_H \left( {\frac{{\partial S}}{{\partial T_H }}} \right) = T_H \left( {\frac{{\partial S}}{{\partial M}}} \right)\left( {\frac{{\partial T_H }}{{\partial M}}} \right)^{ - 1}  = \frac{{\cal A}}{{\cal B}}.
\end{equation}
According to Eqs. (\ref{eq30}) and (\ref{eq31}), the entropy can be rewritten as
\begin{equation}
\label{eq36}
 S = \int {\frac{{4\pi }}{{D - 3}}\left[ {\frac{{\left( {D - 2} \right)\varpi }}{{16\pi GM}}} \right]^{\frac{1}{{D - 3}}} \left\{ {1 + \sqrt {\frac{{\beta \lambda }}{8}}  + } \right.} \left.\beta {\left\{ {\frac{\lambda }{2} + \frac{{\left( {D - 2} \right)\omega ^2 \hbar ^2 }}{{\left( {D - 3} \right)}}\left[ {\frac{{\left( {D - 2} \right)\varpi }}{{16\pi GM}}} \right]^{\frac{2}{{D - 3}}} } \right\}} \right\}dM. \\
\end{equation}
and the $\mathcal{A}$  and  $\mathcal{B}$ in Eq. (\ref{eq35}) are defined by
\begin{eqnarray}
 {\cal A}& = &2^{2 + \frac{4}{{D - 3}}} \left[ {\frac{G}{{\left( {D - 2} \right)\varpi }}} \right]^{\frac{1}{{D - 3}}} \left( {M\pi } \right)^{1 + \frac{1}{{D - 3}}} \left( {4\hbar ^2 \beta \frac{{\left( {D - 2} \right)}}{{\left( {D - 3} \right)}} - \left[ {\frac{{16\pi GM}}{{\left( {D - 2} \right)\varpi }}} \right]^{\frac{2}{{D - 3}}} \left( {4 - \sqrt {2\beta \lambda }  - 2\beta \lambda } \right)} \right)
 \nonumber \\
 \nonumber \\
&\times& \left[ {\left( {1 + \frac{{\sqrt {2\beta \lambda } }}{4} + \frac{{\beta \lambda }}{2}} \right) + \frac{{\hbar ^2 \beta \left( {D - 2} \right)^{1 + \frac{2}{{D - 3}}} }}{{D - 3}}\left( {\frac{{16\pi GM}}{\varpi }} \right)^{\frac{2}{{D - 3}}} } \right],
\label{eq37}
\end{eqnarray}
\begin{equation}
\label{eq38}
 {\cal B} =  -\left\{ {\frac{{12\hbar ^2 \beta \left( {D - 2} \right)}}{{D - 3}} + \left( {2^{\frac{{D + 3}}{{D - 3}}}  - 3 \times 2^{\frac{6}{{D - 3}}} } \right)} \right. \left. { \left[ {\frac{{2GM\pi }}{{\left( {D - 2} \right)\varpi }}} \right]^{\frac{2}{{D - 3}}} \left( {4 - \sqrt {2\beta \lambda }  - 2\beta \lambda } \right)} \right\}. \\
\end{equation}
Assuming $\beta=0$, one obtains the original specific heat of the ST black hole from Eq. (\ref{eq35}). We find that the specific heat  goes to zero at $
M  = \frac{{\left( {D - 2} \right)\varpi _{D - 2} }}{{16\pi G}}\left[ {\frac{{4\left( {D - 2} \right)\hbar ^2 \beta }}{{\left( {D - 3} \right)\left( {4 - 2\beta \lambda  - \sqrt {2\beta \lambda } } \right)}}} \right]^{\frac{{D - 3}}{2}}
$, which is equal to  $M_{min}$ from Eq. (\ref{eq34}). The behaviors of the heat capacity of ST black hole for  $D=4,5$, and $6$  are shown in Fig. 2.
\begin{figure}
\centering
\subfigure{
\begin{minipage}[b]{0.3\textwidth}
\includegraphics[width=1\textwidth]{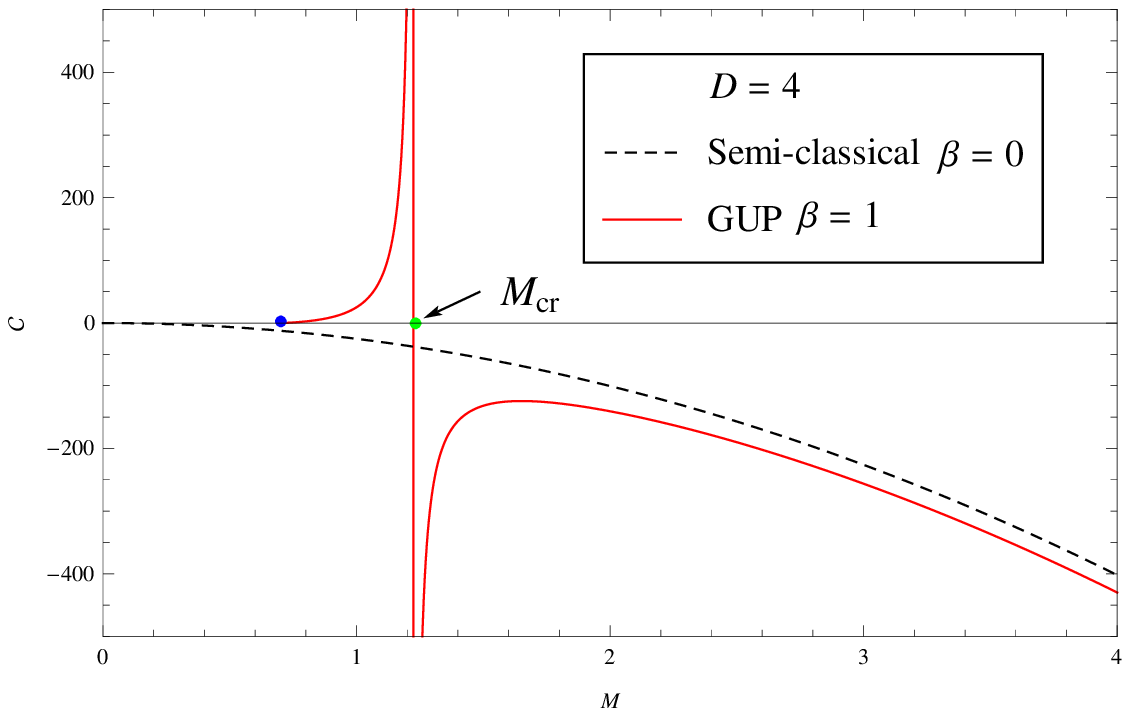}
\end{minipage}
}
\subfigure{
\begin{minipage}[b]{0.3\textwidth}
\includegraphics[width=1\textwidth]{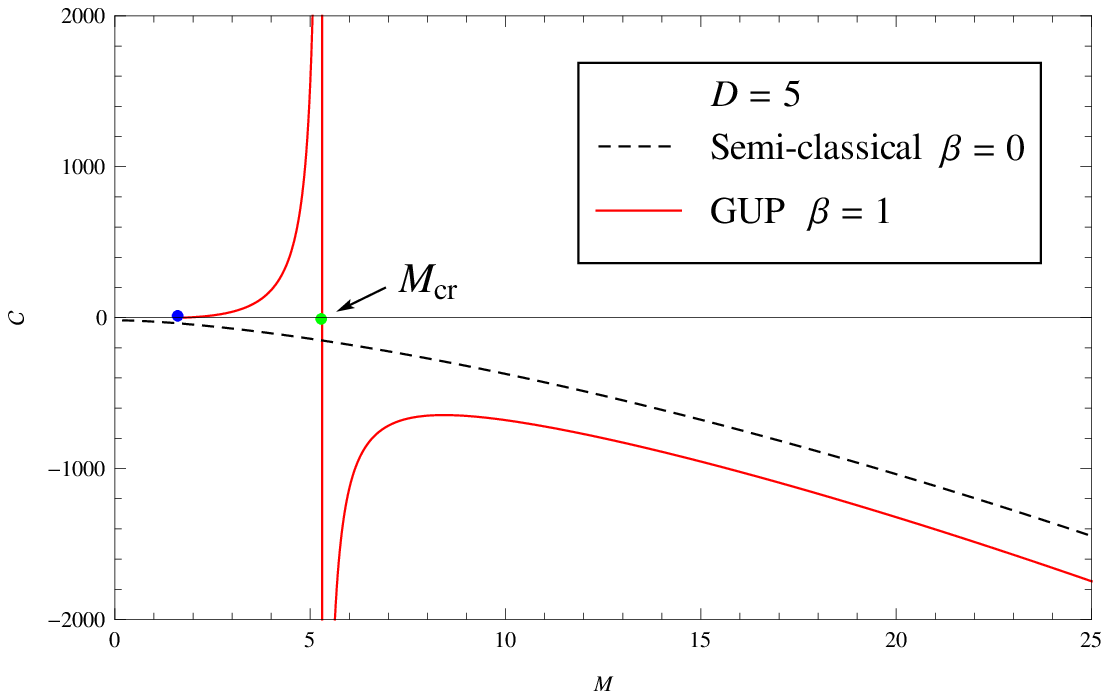}
\end{minipage}
}
\subfigure{
\begin{minipage}[b]{0.3\textwidth}
\includegraphics[width=1\textwidth]{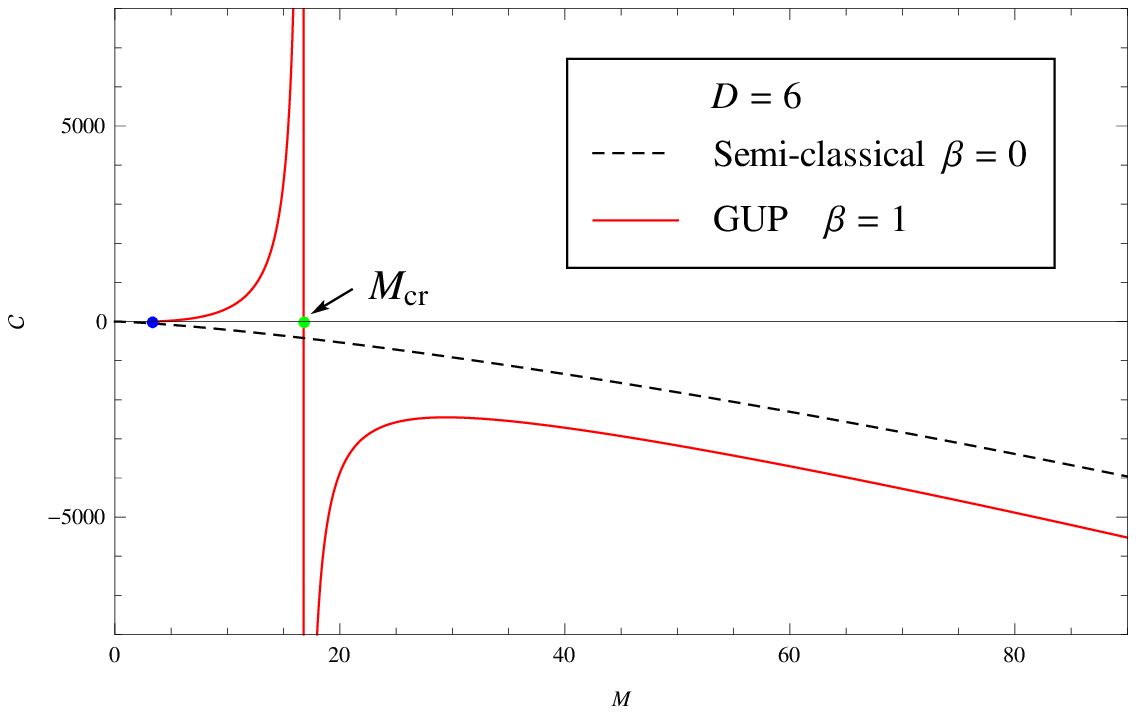}
\end{minipage}
}
\caption{Semi-classical and GUP corrected specific heat of  the ST black hole for different values of  the mass. We assumed $M_P  = c = \hbar = 1$,  $D=4, 5$, and  $6$.}
\label{fig2}       
\end{figure}

In Fig. \ref{fig2}, one can see the specific heat versus the mass of the ST black hole.
Notably, the different values of  $D$ give similar behavior of specific heat.
The black dashed lines correspond to the original specific heat, there are negative values going to zero when  $M\rightarrow0$.
The GUP corrected specific heat is represented by red solid lines. It is clear that the GUP corrected specific heat diverges at the green dot,
where the GUP corrected temperature reaches its maximum value $M_{cr}$. When the mass of the black hole is large enough, the behavior of  the GUP corrected specific heat is similar as the original case. By decreasing the mass of the ST black hole, the GUP corrected specific heat becomes smaller and
departs from the original ST black hole behavior. However, at $M = M_{cr}$, the GUP corrected specific heat has a vertical asymptote at a certain  location; it implies a thermodynamic phase transition occurred  from  $\mathcal{C}<0$ (unstable phase) to  $\mathcal{C}>0$ (stable phase),
and phase transition is also found in the GUP black holes \cite{ch1n+} and the framework of gravity's rainbow \cite{ch8,ch9,ch10}.
 Finally, the GUP corrected specific heat decreases to the zero as mass decreases to  $M_{min}$ (blue dot).
 The $\mathcal{C}=0$  means that the black hole cannot exchange its energy with environment,
 hence the GUP stops the evolution of black holes at this point and leads to the black hole remnant, that is, $M_{\min }  = M_{res}$.

\section{Black hole remnants in the colliders}
The production of black holes at the colliders such as LHC is one of the most exciting predictions of physics.
Due to the Eq. (\ref{eq34}), one can calculate whether the black holes could be formed at the LHC.
The minimum energy needed to form a black hole in a collider is given by
\begin{equation}
\label{eq377}
E_{\min }^{GUP}  = \frac{{\left( {D - 2} \right)M_p }}{{8\Gamma \left( {\frac{{D - 1}}{2}} \right)}}\left[ {\frac{{4\pi \beta _0 \hbar ^2 \left( {D - 2} \right)}}{{c^2 \left( {D - 3} \right)\left( {4 - \frac{{2\lambda \beta _0 }}{{M_p^2 c^2 }} - \sqrt {\frac{{2\lambda \beta _0 }}{{M_p^2 c^2 }}} } \right)}}} \right]^{\frac{{D - 3}}{2}}.
 \end{equation}
In order to investigate the minimal energy for black hole formation,
we use the latest observed limits on the ADD model \cite{ch2c} parameter $M_p$ with a next-to-leading-order (NLO) $K$-factor \cite{ch34d+,ch345d+}.
When setting $\beta_0=c=\hbar=1$ and $\lambda=0.001$, the minimum energy to a form black hole, $E_{\min }^{{{\rm{GUP}}}}$ is shown in Table \ref{tab1}.
\begin{table}[htbp]
\caption {\label{tab1} The latest experimental limits on $M_p$, the minimal energy for black hole formation $E^{\rm GUP}_{\min }$ and $E_{\min }^{{\rm{GR}}}$
in different dimensions $D$.}
\centering
\begin{tabular}{c c c c c}
\toprule
$D$ &  $M_p$ (TeV) &   $E^{\rm GUP}_{\min }$ (TeV) &  $E_{\min }^{{\rm{GR}}}$ (TeV) \cite{ch345d+}\\
\midrule
6&  4.54 &  14.6 &  9.5  \\
7&  3.51 &  17.0 &  10.8 \\
8&  2.98 &  18.7 &  11.8 \\
9&  2.71 &  19.7 &  12.3 \\
10& 2.51 &  19.2 &  11.9 \\
\bottomrule
\end{tabular}
\end{table}

We also compare our results with the results obtained in  the  theory  of Gravity's Rainbow (GR)
 $E_{\min }^{{\rm{GR}}}  = \frac{{\left( {D - 2} \right)}}{{8\Gamma \left( {\frac{{D - 1}}{2}} \right)}}\pi ^{\frac{{D - 3}}{2}} \eta ^{\frac{{D - 3}}{n}} M_p$,
 where $\eta (=1)$ and $n (=2)$ represent rainbow parameter and an integer, $M_p$ is the Planck mass \cite{ch345d+}.
 It is shown that our results are higher than $E_{\min }^{{\rm{GR}}}$. This difference is caused by different modified gravity theories. Quite recently, the protons collided in the LHC have reached the new energy  regime at 13 TeV \cite{ch345f+}, but it is still smaller than the $E^{\rm GUP}_{\min }$ in $D=6$, which implies the black hole cannot be produced in the LHC. This may explain the absence of black holes in current LHC.

 Moreover, we only fix $\beta_0=1$ in Table \ref{tab1}. However, from the expression of $E_{\min }$, we find it is closely related to the dimensionless constant $\beta_0$, which indicates that the different values of $\beta_0$ may lead to different values of the minimum energy for black hole formation. The lower bound of $\beta_0$ can be studied by the following formula:
\begin{equation}
\label{eq39}
 \beta _0  > 4\chi  - \frac{{2\chi ^3 \lambda ^2 }}{{\left( {c^2 M_p^2  + 2\lambda \chi } \right)^2 }} - \frac{{7\lambda \chi ^2 }}{{c^2 M_p^2  + 2\lambda \chi }} - \frac{{c^2 M_p^2 \chi \sqrt {\chi \lambda \left( {8c^2 M_p^2  + 17\chi \lambda } \right)} }}{{\left( {c^2 M_p^2  + 2\chi \lambda } \right)^2 }}, \\
\end{equation}
where $\chi  = \frac{{c^2 \left( {D - 3} \right)}}{{4\pi \hbar ^2 \left( {D - 2} \right)}}\left[ {\frac{{13{\rm{TeV}} \times 8\Gamma \left( {\frac{{D - 1}}{2}} \right)}}{{\left( {D - 2} \right)M_p }}} \right]^{\frac{2}{{D - 3}}}$. The bounds on $\beta_0$ for $D=6,7,8,9,10$ are given in Tab. \ref{tab2}. Combining our results with earlier versions of GUP and some phenomenological implications in \cite{ch14,ch5c,ch7c,ch8c}, it indicates that  $\beta_0\sim1$.
\begin{table}[htbp]
\caption {\label{tab2}The lower bounds on $\beta_0$ for different $D$, we set $c=\hbar=1$ and $\lambda=0.001$.}
\centering
\begin{tabular}{c c c c c c}
\toprule
$D$ &  $6$  &   $7$ &   $8$&   $9$ &   $10$\\
\midrule
$\beta_0$&  0.9216&  0.8740&  0.8642&   0.8707&   0.8944\\
\bottomrule
\end{tabular}
\end{table}

\section{Conclusions}
In this work, we have investigated the GUP effect on the thermodynamics of ST black hole. First of all, we derived the modified Hamilton-Jacobi equation
 by employing the GUP with a quadratic term in momentum. With the help of the modified Hamilton-Jacobi equation,
 the quantum tunneling from the ST black hole has been studied. Finally, we obtained the GUP corrected Hawking temperature,
 entropy, and heat capacity. For the original Hawing radiation, the Hawking temperature of the ST black hole is related to its mass. However, our results showed that if the effect of quantum gravity is considered, the behavior of the tunneling particle on the event is different from the original case, and the GUP corrected thermodynamic quantities are not only sensitively dependent on the mass $M$ and the spacetime dimension $D$ of ST black hole, but also on the angular momentum parameter $\lambda$ and the quantum gravity term  $\beta$. Besides, we found that the GUP corrected Hawking temperature is smaller than the original case;  it goes to zero when the mass of ST black hole reaches the minimal value $M_{min}$, which is of the order of the Planck scale,
and it predicts the existence of a black hole remnant. For confirming the black hole remnant, the GUP corrected heat capacity has also been analyzed.
   It was shown that the GUP corrected heat capacity has a phase transition at  $M_{cr}$, where the GUP corrected temperature reaches its maximum value;
   then the GUP corrected heat vanishes when the mass approaches to  $M_{min}$ in the final stages of black hole evaporation.
   At this point, the ST black hole does not exchange the energy with the environment, hence the remnant of ST black hole is produced.
   The reason for this remnant is related to the fact that the quantum gravity effect is running as the size of the black hole approaches to the Planck scale.
   The existence of a black hole remnant implies that black holes would not evaporate, its information and singularity are enclosed in the event horizon.
   Finally,  we discussed the minimum energy to form black hole in the LHC.
   The results showed that the minimum energy to form black hole in our work is larger than the current energy scales of LHC,
   this may explain why one cannot observe a black hole in the LHC.
   Our results are support by the results obtained in the framework of gravity's rainbow \cite{ch8,ch9,ch10}.
   Therefore, we think that the GUP effect can effectively prevent the black hole from evaporating completely, and this  may solve the information loss
   and naked singularities problems of black holes \cite{ch35,ch36}.

\vspace*{3.0ex}
{\bf Acknowledgements}
\vspace*{1.0ex}

This work is supported by the Natural Science Foundation of China (Grant No. 11573022).

\end{document}